\documentclass[
 reprint,
 amsmath,amssymb,
 aps
]{revtex4-1}

\usepackage{graphicx}
\usepackage{dcolumn}
\usepackage{bm}
\usepackage{amsmath,amssymb,graphicx,bbm,mathrsfs,nicefrac}
\usepackage{tikz}
\usepackage{amsmath}
\usepackage{amssymb}
\usepackage{graphicx,textcomp,float,gensymb,wrapfig, enumitem,comment,dsfont,framed,slashed,appendix,wrapfig,xcolor}
\usepackage{ bbold }
\usepackage{braket} 
\usepackage{ mathrsfs }
\usepackage[small]{caption}
\usepackage{subcaption}
\usepackage{etoolbox,hyperref,makecell}
\usepackage{feynmp}
\usetikzlibrary{arrows.meta}
\usepackage{empheq}
\newcommand{\bea}{\begin{eqnarray}}  
\newcommand{\eea}{\end{eqnarray}}

\usepackage[utf8]{inputenc}

\begin{document}

\title{
How to Discover Mirror Stars
}

\author{David Curtin}
 \email{dcurtin@physics.utoronto.ca}
\author{Jack Setford}%
 \email{jsetford@physics.utoronto.ca}
\affiliation{
Department of Physics, University of Toronto, Canada
}

\date{\today}

\begin{abstract}

Non-minimal hidden sectors are an important generic possibility and arise in highly motivated theories like Neutral Naturalness. 
A fraction of dark matter could therefore have hidden interactions analogous to Standard Matter (SM) electromagnetism and nuclear physics. 
This leads to the formation of  \emph{Mirror Stars}: dark-sector analogues of regular stars that shine in dark photons. 
We examine the visible signatures of Mirror Stars in observations for the first time. 
If the dark and visible photon have a small kinetic mixing, SM matter is captured in Mirror Star cores, giving rise to an optical signal similar to but much fainter than white dwarfs, as well as a separate X-ray signal that represents a direct window into the Mirror Star core.
This robust and highly distinctive signature is a smoking gun of Mirror Stars and could be discovered in  optical and X-ray searches.

\end{abstract}

\pacs{Valid PACS appear here}
\maketitle

\textbf{1.~Introduction ---} The Standard Model (SM) is a highly non-minimal theory, with many different particle species, interactions, and mass scales giving rise to the complexity of our visible universe. 
All or part of Dark Matter (DM) may be similarly complicated, and indeed this possibility arises in a variety of fundamentally motivated models~\cite{Chacko:2005pe, Chacko:2018vss, Garcia:2015loa, Craig:2015xla, Garcia:2015toa, Farina:2015uea, Cheng:2018vaj, Hochberg:2018vdo,
Zurek:2013wia,Petraki:2013wwa}. It is hence vital to study the signatures of Dark Complexity.

Recent years saw significant progress in the study of next-to-minimal examples, like DM with additional interactions or a small number of states~\cite{Zurek:2013wia,Petraki:2013wwa, 
Alexander:2016aln,Battaglieri:2017aum,
Krnjaic:2017tio,Tulin:2017ara,Gluscevic:2019yal,DeRocco:2019njg,Gresham:2018rqo,Chang:2018rso,Fan:2013yva,Fan:2013tia,McCullough:2013jma,Agrawal:2016quu,Agrawal:2017rvu,Agrawal:2017pnb,
Acevedo:2019gre,Bhoonah:2018gjb,Bramante:2016rdh}, but these approaches are limited in the variety of phenomena they can explore. 
On the other hand, the study of more complicated dark sectors~\cite{Chacko:2018vss, Berlin:2018tvf,Grossman:2017qzw,Kuflik:2015isi,Hochberg:2014dra,Brax:2019koq,Kribs:2018ilo,Cheng:2018vaj,Renner:2018fhh,Garcia:2015loa, Craig:2015xla, Garcia:2015toa, Farina:2015uea, Cheng:2018vaj,Hochberg:2018vdo,
Zurek:2013wia,Petraki:2013wwa,
Khlopov:1989fj,Foot:1999hm,Foot:2000vy,Berezhiani:2003xm,Foot:2004pa,Berezhiani:2005ek,Foot:2014mia,Michaely:2019xpz,
Detmold:2014qqa,Detmold:2014kba,Krnjaic:2014xza} is made daunting by the sheer multitude of possibilities
and the difficulty of making physical predictions. Our approach will be
the study of dark sectors that are related (but not identical) to the SM by some symmetry.
This allows us to derive physical predictions with guidance from known physics, while providing a starting point to understand the signals of true Dark Complexity. 

A suitable and highly motivated benchmark model is the asymmetrically reheated incarnation \cite{Chacko:2016hvu,Craig:2016lyx} of the Minimal Twin Higgs (MTH)  \cite{Chacko:2005pe}, which solves the little hierarchy problem without colored top partners, and  predicts a mirror sector that is a copy of the SM with a mirror Higgs VEV $v_B$ a few times higher than the visible Higgs VEV $v_A$.
The rich cosmological signatures of this scenario, including effects of an asymmetric relic density of mirror baryons on Large Scale Structure, were recently explored~\cite{Chacko:2018vss}.
The similarities between SM matter and mirror matter also make it clear that the mirror matter could cool and clump in our galaxy~\cite{MTHastro},  leading to the formation of \emph{Mirror Stars} (MS) that fuse mirror nuclei and shine in mirror photons. 

The possibility that some fraction of DM could form Mirror Stars is  both extremely intriguing and quite general for a complex dark sector, requiring only a massless dark photon (along with e.g. an early temperature asymmetry between the dark and visible sectors to avoid CMB constraints \cite{Aghanim:2018eyx}) and some processes analogous to nuclear physics. 
The idea of Mirror Stars has been discussed in the context of Mirror-DM models~\cite{Foot:1999hm, Foot:2000vy}
but their more general nature and non-gravitational observational consequences were never carefully explored. 
We demonstrate that Mirror Stars lead to spectacular astrophysical signals if the SM ($A^\mu$) and mirror ($A^\mu_D$)  photons have a tiny kinetic mixing $\frac{1}{2} \epsilon F_{\mu \nu} F_D^{\mu \nu}$ \cite{Holdom:1985ag}, which is expected to exist because it cannot be forbidden by symmetries, is dynamically generated in many UV-completions, and violates no cosmological constraints for $\epsilon \lesssim 10^{-9}$~\cite{Chacko:2018vss, Vogel:2013raa}. Indeed, small values of $\epsilon \sim 10^{-13}-10^{-10}$ are well motivated in the MTH model~\cite{Koren:2019iuv}. 

In this letter we show how to analytically estimate these signals of Mirror Stars, with a more detailed calculation presented in a companion paper~\cite{bigpaper}. 
We find that Mirror Stars capture SM matter in their cores.
This ``SM nugget'' gets heated up to $T \sim 10^4$~K by $\epsilon^2$-suppressed interactions with the mirror matter, giving rise to an optical signal similar to but much fainter than standard white dwarfs.
We also show for the first time that \emph{mirror Thomson conversion} allows thermal dark photons from the Mirror Star core to be converted to visible X-rays that escape the nugget, providing a direct window into the MS interior.
This double signature is extremely distinctive and can be discovered in  optical and X-ray searches~\footnote{This double-signature is also many orders of magnitude larger than the surface luminosity of Mirror Stars in visible photons, which is $\epsilon^2$ $\times$ their hidden photon luminosity.}

The question of how the detailed properties and distribution of Mirror Stars follow from the details of the dark sector is highly non-trivial, and we will explore it in future work. 
In this letter we instead study a simplified scenario where the dark sector, making up a subdominant fraction of DM, contains mirror quarks, leptons, and gauge forces that are perfect copies of their SM analogues. 
Standard stellar evolution codes can then be used to compute the properties of some benchmark Mirror Stars, see Table~\ref{table:benchmarks}. 
This allows us to develop our calculation of the Mirror Star signal, which is then readily applicable to Mirror Stars that arise in theories of Neutral Naturalness \cite{Chacko:2005pe,Burdman:2006tz} or  more exotic dark sectors. 

\newcommand{\SMcapture}{Section 2}
\newcommand{\OpticalDepth}{Section 3}
\newcommand{\Bremsstrahlung}{Section 4}
\newcommand{\Xray}{Section 5}
\newcommand{\Conclusion}{Section 6}

\textbf{2.~SM Baryon Capture in Mirror Stars ---} 
In the presence of a small dark photon mixing, the MS captures SM material from the Interstellar Medium (ISM) that accumulates in its core. We need to determine both the size and the properties of this ``SM nugget''.

Capture proceeds via interactions of ISM SM baryons with the Mirror Star matter (mirror capture) as well as already captured SM matter (self-capture).
The following calculation is very analogous to the problem of dark matter capture in the sun, see e.g. \cite{Gould:1987ir, Zentner:2009is, Catena:2015uha, Petraki:2013wwa}.

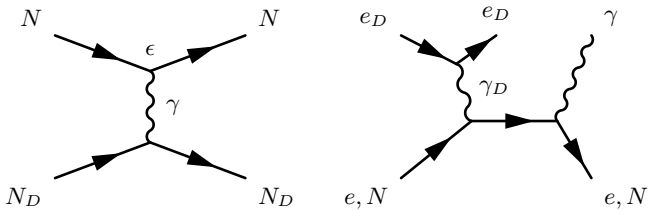
\begin{figure}[t!]
\vspace{5mm}
\centering
\begin{fmffile}{nuclei_scattering_letter}
\begin{tikzpicture}[baseline=(current bounding box.center)]
\node{
\fmfframe(1,1)(1,1){
\begin{fmfgraph*}(90,54)
\fmfleft{i1,i2}
\fmfright{f1,f2}
\fmf{fermion}{i2,b}
\fmf{fermion}{i1,a}
\fmf{photon,label=$\gamma$}{a,b}
\fmf{fermion}{b,f2}
\fmf{fermion}{a,f1}
\fmflabel{$N_D$}{i1}
\fmflabel{$N$}{i2}
\fmflabel{$N_D$}{f1}
\fmflabel{$N$}{f2}
\fmflabel{$\epsilon$}{b}
\end{fmfgraph*}
}
};
\end{tikzpicture}
\end{fmffile}
\hspace{5mm}
\begin{fmffile}{photon_conversion_letter}
\begin{tikzpicture}[baseline=(current bounding box.center)]
\node{
\fmfframe(1,1)(1,1){
\begin{fmfgraph*}(90,54)
\fmfleft{i1,i2}
\fmfright{f1,f2}
\fmftop{f3}
\fmf{fermion}{i2,a}
\fmf{fermion}{i1,b}
\fmf{photon,label=$\gamma_{D}$}{b,a}
\fmf{fermion}{a,f3}
\fmf{fermion}{b,c}
\fmf{fermion,tension=1.5}{c,f1}
\fmf{photon}{c,f2}
\fmflabel{$e,N$}{i1}
\fmflabel{$e_D$}{i2}
\fmflabel{$e,N$}{f1}
\fmflabel{$\gamma$}{f2}
\fmflabel{$e_D$}{f3}
\end{fmfgraph*}
}
};
\end{tikzpicture}
\end{fmffile}
\vspace{3mm}
\caption{Important interactions between between mirror and visible matter. Left: nucleus-nucleus scattering. Right: Thomson conversion of a mirror photon to a SM photon. } 
\label{figure:diagrams}
\end{figure}

Mirror capture is determined by the Rutherford scattering cross section, see Fig.~\ref{figure:diagrams} (left):
\begin{equation}
\label{equation:cross_section}
\frac{d\sigma}{d E_R} = \frac{2\pi\epsilon^2\alpha^2 Z_1^2 Z_2^2}{m_T v^2 E_R^2},
\end{equation}
where $v$ is the relative incoming velocity, $m_T$ is the mass of the target nucleus, and $E_R$ is its recoil energy. 
We can treat the entire ISM as ionized, since incoming atoms falling onto the gravity wells of our benchmark stars are fast enough to make the effect of atomic form factors small. The SM nugget as a whole will always neutralize by attracting electrons from the ISM as needed.
The capture calculation is standard~\cite{Gould:1987ir, Zentner:2009is, Catena:2015uha, Petraki:2013wwa}, for more details see \cite{bigpaper}. A very useful approximation in our scenario is that the Mirror Star escape velocity $v_\mathit{esc}$ (around $600$~km/s for the 1 $M_{sun}$ benchmark) is much larger than the relative velocity $u$ of the ISM to the Mirror Star, which we take to be $u \approx 20$ km/s from the velocity dispersion of stars and gas in our local stellar neighbourhood  \cite{Dehnen:1997cq, LopezSantiago:2006xv}. 
As long as the hierarchy between mirror and SM nuclear masses is less severe than $\frac{1}{4}\left({u}/{v_\mathit{esc}}\right)^2$, the final result for the  mirror capture (denoted by $m$ superscript) is independent of $v_\mathit{esc}$:
\begin{eqnarray}
\label{equation:capture_rate}
\frac{dN_i^{(m)}}{dt} &=& n^\mathit{ISM}_i\sum_j\frac{4\pi c^4 N_j \epsilon^2\alpha^2 Z_i^2 Z_j^2}{m_i m_j u^3} 
\end{eqnarray}
for each incoming species $i$, where $n^\mathit{ISM}_i$ is the local ISM density of the incoming species, $N_j$ is the total number of scattering targets of species $j$, $m_i$ and $Z_{i}$ are the mass and nuclear charge of the relevant species, and the sum is over different target species in the Mirror Star.
The expression for self-capture rate $dN_i^{(s)}/dt$ is analogous, without the $\epsilon^2$ factor.

\begin{table}[t]
\begin{tabular}{| c | c | c | c |}
\hline
$M$ / $M_\mathit{sun}$ & 1 & 5 & 50 \\
\hline
He / H & 0.24 & 0.24 & 0.24 \\
$R$ / $R_\mathit{sun}$ & 1 & 3.80 & 16.1 \\
$T_\mathit{core}$ / $10^7$ K & 1.54 & 2.83 & 4.13 \\
$L$ / $L_\mathit{sun}$ & 0.96 & 721 & $5.18\times10^5$ \\
$\tau_\mathit{star}$ / years & $4.3\times10^9$ & $5.6\times10^7$ & $2.3\times10^6$ \\
$n_\mathit{core}$ / cm$^{-3}$ & $4.5\times10^{25}$ & $6.2\times10^{24}$ & $7.4\times10^{23}$\\
$\epsilon_\mathit{crit}^\mathit{mirror}$ & $2.6\times10^{-9}$ & $5.9\times10^{-9}$ & $1.3\times10^{-8}$\\
$\epsilon_\mathit{crit}^\mathit{self}$ & $2.5\times10^{-16}$ & $3.0\times10^{-15}$ & $8.2\times10^{-14}$\\
\hline
\end{tabular}
\caption{Properties of the three benchmark Mirror Stars, computed in MESA~\cite{Paxton2011, Paxton2013, Paxton2015, Paxton2018, Paxton2019}: mass, helium mass fraction, radius, core temperature, luminosity, age of star (half main sequence lifetime), total number of atoms, atomic number density at the core, critical values of $\epsilon$ above which the mirror and self capture rates are geometric.
}
\label{table:benchmarks}
\end{table}

We consider capture of hydrogen and helium only, and assume that their average densities over the path the Mirror Star has traced are given by $n_H^\mathit{ISM} = 1\,\textrm{cm}^{-3}$ and $n_{He}^\mathit{ISM} = 0.1\,\textrm{cm}^{-3}$, roughly in accordance with the average values for our galaxy \citep{2011piim.book.....D}. It is clear that the total amount captured simply scales with $n_i^\mathrm{ISM} \epsilon^2/u^3$, where strictly speaking $1/u^3$ should be averaged over the ISM velocity distribution, but we use 20 km/s for simplicity.

If the capture rate is so large that every incoming particle is captured, capture become geometric, i.e. the cross section is given by the physical size of the target. The geometric capture rate is independent of the scattering cross section, and is given approximately by \cite{Petraki:2013wwa}
\begin{equation}
\frac{dN_i^{(geo)}}{dt} = n_i^\mathit{ISM}\sqrt\frac{3}{2} \frac{\overline v_	\mathit{esc}^2}{u} \pi R^2,
\end{equation}
where 
 $R$ is the radius of the capture region (size of Mirror Star/SM nugget for mirror/self capture)
 and
 $\overline v_\mathit{esc}$ is the escape velocity (averaged over the Mirror Star/SM nugget for mirror/self capture).
 For our benchmark Mirror Stars and range of $\epsilon$, self-capture becomes geometric very quickly in much less than 10\% of the stellar lifetime, making it $\epsilon$-independent, while mirror capture is never geometric, scaling with $\epsilon^2$, see Table~\ref{table:benchmarks}. Self-capture becomes comparable to mirror capture for $\epsilon\sim10^{-11}$ in all our benchmarks.
  
 The radius of the SM nugget can be estimated from the virial theorem, equating its gravitational potential energy with its thermal kinetic energy:
\begin{equation}
\label{equation:virial}
R_\mathit{nugget} = r_\mathit{virial} = \left( \frac{9 T_\mathit{SM}}{4\pi G \rho_\mathit{mirror} \overline\mu_\mathit{SM}} \right)^{1/2}.
\end{equation}
$T_\mathit{SM}$ is the average temperature of the nugget, $\rho_\mathit{mirror}$ is the density of mirror matter at the core of the MS, and $\overline\mu_\mathit{SM}$ is the average mass of captured SM particles.
We therefore need to know the temperature of the captured matter to know the geometric self-capture rate. 
As we show in \Bremsstrahlung, this temperature is always in the range of $4000-7000$~K for the benchmark stars we consider.
This temperature is much lower than $T_{core}$ due to the $\epsilon^2$ suppression of heat transfer compared to the unsuppressed cooling by bremsstrahlung emission, and is in the range $\lesssim 10^4$~K since that is where ionization and hence cooling efficiency increases sharply with temperature. 
The modest dependence of $r_{virial}$ on $T_\mathit{SM}$ makes $T_\mathit{SM} \approx 6000$K is a good ansatz to estimate the self-capture rate. 
It is easy to solve consistently for the capture rate and temperature, but our ansatz is sufficient to estimate signal to better than a factor of 2.

With the capture rate determined, we assume that the total amount of captured material is, on average, given by simply multiplying that rate by half the MS main sequence lifetime. 
Note that we ignored evaporation of SM matter from the nugget in this simple estimate. In~\cite{bigpaper} we show that this is a valid assumption, since only freshly captured H or He nuclei  have enough velocity to be kicked out of the gravity well by collisions with thermal mirror ions, but such collisions have such low probability that most SM nuclei settle into the nugget before evaporating.

\textbf{3.~Optical Depth ---} 
The properties of the SM nugget depend on its optical depth to SM photons in three frequency ranges, which in turn depend on the density and temperature of the nugget. 

The nugget is optically thick to photons with the correct energy to ionize atoms~\cite{1986rpa..book.....R}. This means that we can use Saha's equation to compute the ionization of the SM nugget, and that cooling via inter-atomic collisional processes can be neglected.

For most of the cases we consider, the nugget is optically thin for photons far below the ionization threshold, allowing it to cool via bremsstrahlung processes. 
For higher densities and temperatures closer to $10^4$ K, the larger degree of ionization means the nugget can become optically thick due to free-free absorption \cite{2011piim.book.....D}. 

In that case the nugget cools as a black body from surface emission. We discuss this in more detail in~\cite{bigpaper}, but the final expected MS signal will not differ greatly from the optically thin case, since the total luminosity, determined by the heating rate, remains the same. 

The optical depth of the nugget to X-ray frequencies far above the ionization threshold is relevant to determine the size of the X-ray signal. X-rays will scatter (almost) elastically from atoms and free charges alike, and can be absorbed in photoionization. We discuss this in \Xray~when calculating the X-ray signal.

Finally, in order for the SM nugget to be observable, the Mirror Star  has to be optically thin to SM photons. This is certainly the case for most benchmarks we consider, though there is some attenuation of the optical signal at low frequencies due to free-free absorption in the fully ionized MS mirror matter for $\epsilon \gtrsim 10^{-10}$. For details, see~\cite{bigpaper}. We neglect this effect in the present analysis for simplicity since it does not affect our conclusions.

\textbf{4.~Bremsstrahlung Cooling Signal ---} 
We assume that the SM nugget is a sphere of constant density, with radius given by Eqn.~\eqref{equation:virial}.
The nugget is heated via collisions of SM atoms (the ionized fraction will be very small) with mirror ions in the core. Since the nugget is much colder than the core, we can treat the SM atoms as being at rest. 
The relative velocity with mirror ions $v_{rel}$ is sampled from the thermal distribution of the mirror nuclei, but a good approximation is to set it to the thermal average  $\sqrt{3 T_\mathit{mirror}/\overline m_\mathit{mirror}}$, with  $\overline m_\mathit{mirror}$ being the average mirror nuclear mass. 
We then compute the average energy transfer given the momentum dependence of the ion-atom scattering cross section, which is given by Eqn.~\ref{equation:cross_section} with the replacement
 $E_R \to E_R + (2 m_T a_0)^{-2}$~\cite{Cline:2012is}:
 \begin{eqnarray}
\label{equation:heating_rate}
\frac{dP^i_\mathit{coll}}{dV} &=& n_\mathit{mirror} n_\mathit{SM} v_\mathit{rel} \int_{0}^{E_R^\mathit{max}} dE_R \, E_R \, \frac{d\sigma}{dE_R} 
\\
\nonumber 
&\approx& n^i_{mirror} n_{SM} \frac{2\pi\epsilon^2\alpha^2 Z_{SM}^2 Z_i^2}{m_{SM} v_{rel}} \left(\log\frac{8 \mu^2 v_{rel}^2}{(1/a_0)^2} - 1\right),
\end{eqnarray}
where $a_0$ is the SM Bohr radius,
SM stands for SM H or He, 
$\mu$ is the reduced mass of the atom and colliding mirror ion, and $E_R^{max} = 4 \mu^2 v_{rel}^2 /m_{SM}$.
 Since the nugget is much smaller than the Mirror Star, $n_{mirror}$  and $T_{mirror}$ can be taken to be their values at $r = 0$, so the total collisional heating rate is trivially obtained by $n_{SM} \to N_{SM}$. 

Conversion of mirror photons into SM X-ray photons, discussed in \Xray, can be a source of heating if the X-rays are absorbed by the nugget before escaping. However, X-ray heating is always subdominant to collisional heating, so we can neglect it in our estimates.

\begin{figure}[t]
\includegraphics[scale=0.35]{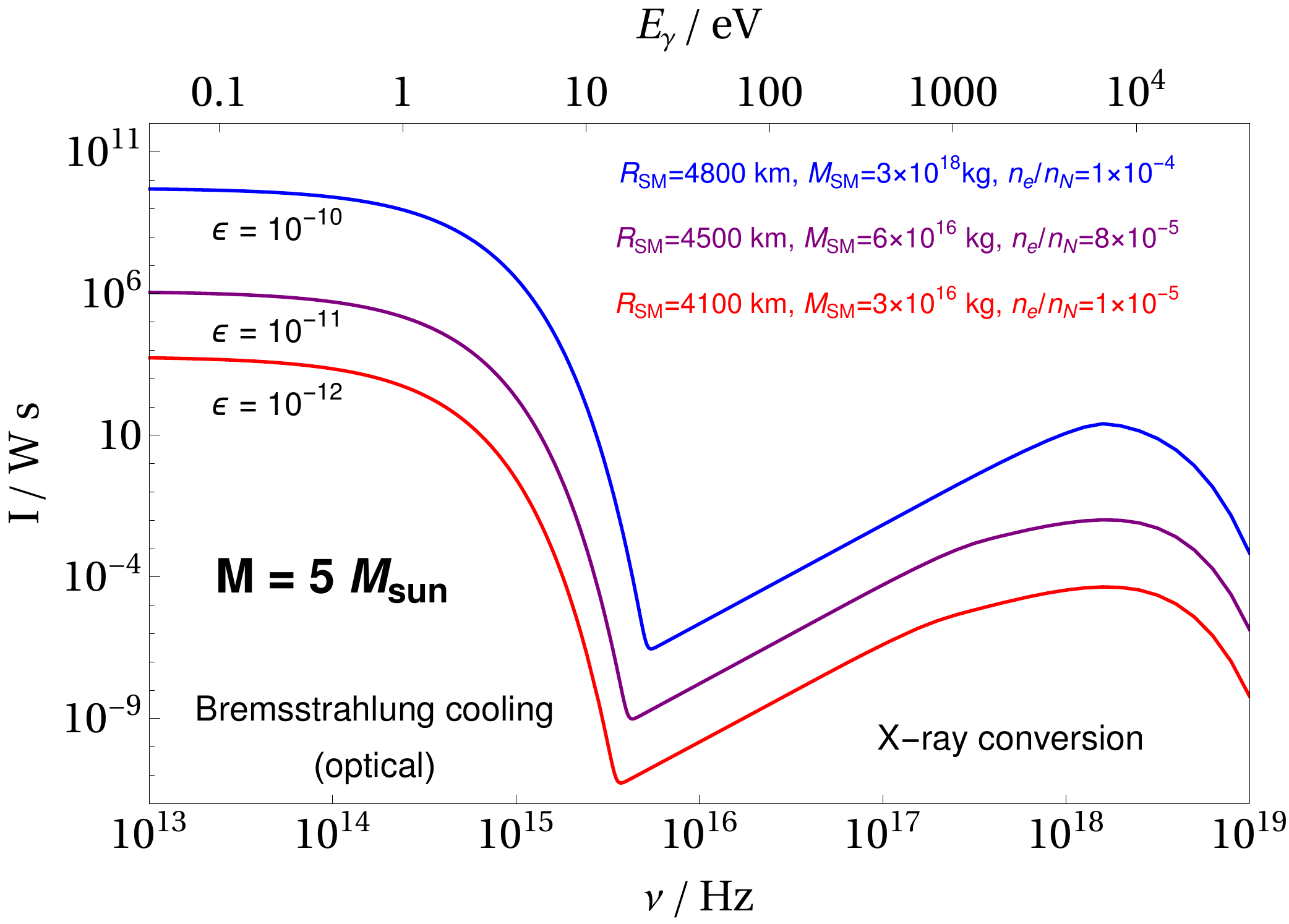}
\caption{Estimated emission spectra for a 5 solar mass benchmark star for three different values of $\epsilon$. The X-ray peak is closely related to the Mirror Star core temperature.
}
\label{figure:spectra}
\end{figure}

The collisional heating rate, and hence the total luminosity of the nugget in optical/thermal frequencies, is independent of the temperature of the nugget. However, the shape of the emission spectrum does depend on temperature. 
We solve for  $T_\mathit{SM}$, by equating the heating power to the total cooling power assuming bremsstrahlung cooling of an optically thin medium~\cite{2011piim.book.....D},
\begin{equation}
\frac{dP_{ff}}{dV} = \frac{16}{3} \left( \frac{2\pi}{3} \right)^{1/2} \frac{\alpha^3}{m_e^2} \,(m_e T_\mathit{SM})^{1/2} \,\langle g_{ff} \rangle_T\, Z_i^2\, n_e\, n_i,
\end{equation}
where $\langle g_{ff} \rangle_T$ is the frequency averaged free-free Gaunt factor at temperature $T$ (close to unity in our case). 
Self-consistently assuming the ionization to be small, the solution to Saha's equation for pure hydrogen (similarly with helium) takes the simple form
\begin{equation}
n_e n_i = n_{SM} \left(m_e T_\mathit{SM}/2\pi\right)^{3/2} \exp\left(-\frac{\omega_0}{T_\mathit{SM}} \right),
\end{equation}
where $\omega_0$ is the ionization energy. We see that $n_e n_i \propto n_\mathit{SM}$, which drops out when equating heating and cooling rates. The equilibrium temperature of the SM nugget is therefore independent of density, and our approximation of constant nugget density drops out of the temperature calculation. 
Solving for $T_\mathit{SM}$ as a function of $\epsilon$ for each MS, we verify our original claim that the temperature of the nugget lies in the range $4000-7000$ K.

In Figure~\ref{figure:spectra} we plot the emission spectra of the nugget for different values of $\epsilon$~\footnote{Note that for $\epsilon = 10^{-10}$, the neglected effects of low-frequency attenuation in the MS and optical thickness of the SM nugget change some details~\cite{bigpaper}, but this does not change our conclusions.}. The shape of the spectrum is flat for low frequencies, which is characteristic of bremsstrahlung emission \citep{2011piim.book.....D}. 

\textbf{5.~X-ray Conversion signal ---} 
Mirror photons in the MS core can undergo elastic ``Thomson conversion'' to a SM photon off electrons or nuclei in the nugget, $\gamma_D + e/N \to \gamma + e/N$, see Fig.~\ref{figure:diagrams} (right).
These X-rays can escape the nugget and lead to an observable signal that \emph{directly} probes the Mirror Star interior, revealing for example the core temperature. 
As far as we are aware, this is the first time this conversion signature has been discussed in the context of Mirror Stars.

For $T_\mathit{core}$ far above SM atomic binding energies, Thomson conversion proceeds identically for both bound and free SM electrons/nuclei. 
The conversion rate depends on the Thomson scattering cross section, which is frequency independent. Therefore the spectrum of converted X-rays will match the mirror photon spectrum in the core, well approximated by a black body with temperature $T_\mathit{core}$. The conversion power per unit volume is
\begin{equation}
\frac{d P_\mathit{conv}}{d V\,d \nu} = \epsilon^2 n_{SM}\sigma_\mathit{thoms} 4\pi B_\nu(\nu, T_\mathit{core}),
\end{equation}
where $B_\nu$ is the Planck spectral radiance function for a black body and $\sigma_\mathit{thoms}$ is the Thomson scattering cross section. This gives the familiar $T_\mathit{core}^4$ dependence when integrated over all frequencies.

\begin{figure}%
\includegraphics[scale=0.38]{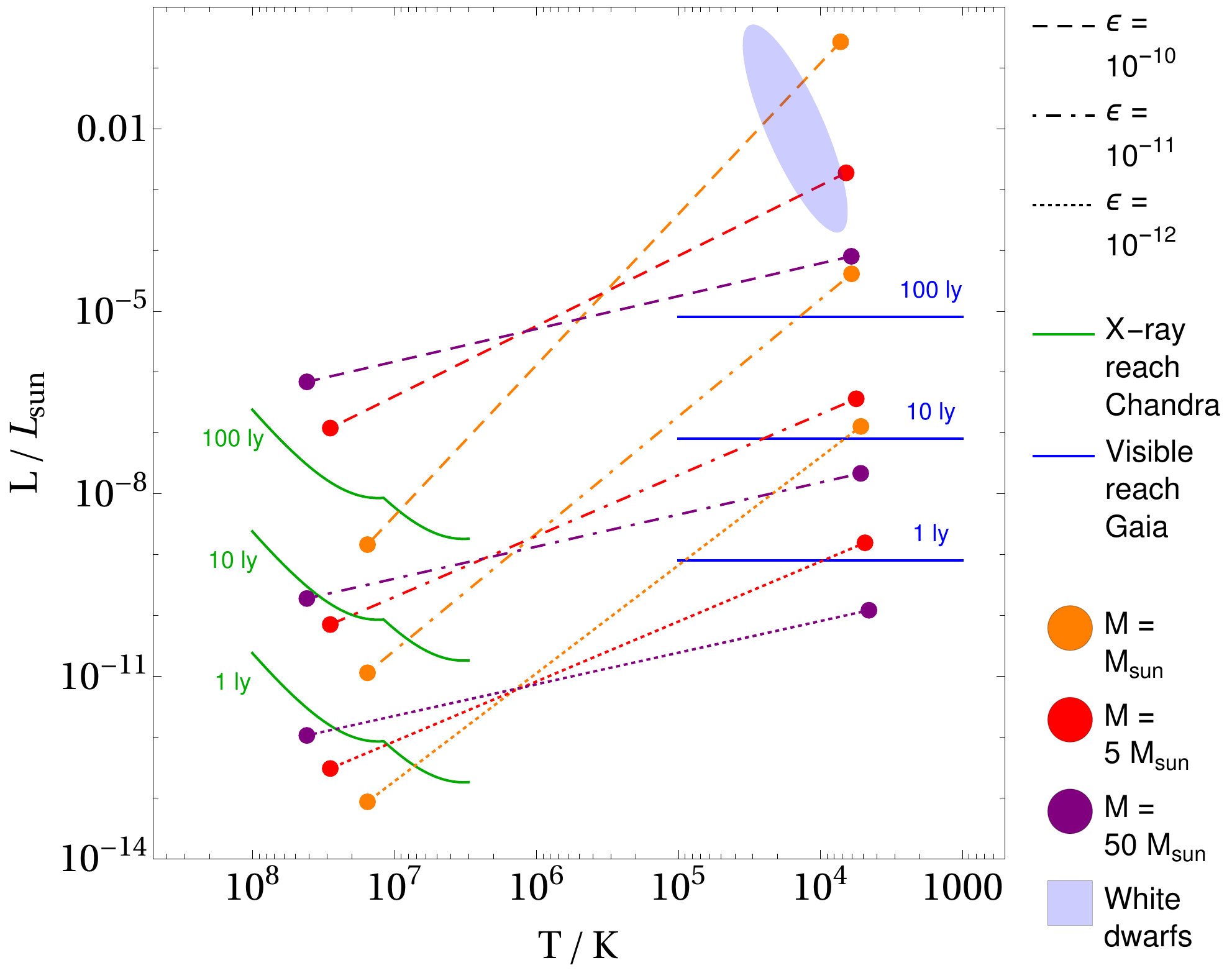}
\caption{A Hertzsprung-Russell diagram showing the dual signatures of our Mirror Star benchmarks estimated in this letter. Each MS is represented by two points connected by a line. The solid lines show the approximate distances up to which such objects could be observed in Gaia or Chandra.}
\label{figure:HRplot}
\end{figure}

Converted X-rays have a chance of ionizing neutral atoms and being absorbed. They also lose on average $\Delta E_\gamma = E_\gamma^2/m$ worth of energy in each scatter with a free electron or nucleus of mass $m$.  In computing the X-ray spectrum emitted by the nugget we have to account for both absorption and energy loss.
We do not solve a full diffusion equation for the escaping X-ray photons, but we can estimate the total escaping X-ray power as follows:
\begin{multline}
\label{equation:xrays}
\frac{dP_\mathit{x\textrm{-}ray}}{d\nu_\mathit{obs}}=\int_0^{R_\mathit{nugget}}dr \,4\pi r^2 \int_0^\infty d\nu_i\,\frac{\nu_f}{\nu_i}\frac{dP_\mathit{conv}}{dV \,d\nu_i} \times \\
\Theta\big(\lambda_\mathit{abs}(\nu_f)-N_\mathit{scat}(r)\lambda_\mathit{scatter}\big) \delta\big( \nu_\mathit{obs} - \nu_f(\nu_i, r) \big).
\end{multline}
In this expression, $\nu_i$ represents the frequency of a photon when it first converts, while $\nu_f$ is its energy after $N_\mathit{scat}$  scatters: $\nu_f = \nu_i m/(\nu_i N_\mathit{scat} + m)$.
$\lambda_{abs, scatter}$ are the mean free paths for absorption or Thomson scattering computed at the starting point.
 The Heaviside function accounts for absorption, and ensures that photons do not contribute to the signal if they must travel further than the absorption path length  before escaping the nugget, the absorption path length being a function of frequency. The number of scatters required for the photon converting at position $r$ to escape is estimated by assuming the photon must random walk a distance $R_\mathit{nugget} - r$, leading to $N_\mathit{scat} = (R_\mathit{nugget} - r)^2/\lambda_\mathit{scatter}^2$.
 Assuming the nugget has a constant density profile yields the X-ray contributions to the emission spectra of Figure~\ref{figure:spectra}, agreeing to better than a factor of 2 with a more realistic profile~\cite{bigpaper}.

\textbf{6.~Discussion and Conclusions ---} 
We summarise our results in Figure~\ref{figure:HRplot}, plotting the luminosity and characteristic temperature of both the thermal and converted X-ray Mirror Star signatures on a Hertzsprung-Russell diagram.
Mirror Stars look spectacularly alien compared to standard astrophysical expectations: the optical signal is similar to that of a white dwarf but typically orders of magnitude too faint in absolute luminosity to be compatible with its high temperature. This could be discovered in full-sky surveys like Gaia~\cite{2016A&A...595A...1G} out to the distances indicated by blue lines (which also provides a parallax measurement to determine absolute luminosity).
The discovery of such faint Mirror Star candidates would prompt extremely detailed study with an X-ray observatory:
Chandra could see the X-ray signal  
roughly out to distances indicated by green lines 
with an exposure equal to the Hubble Deep Field North~\cite{Brandt:2001vb}.
Detection of this black-body-like X-ray signal would be a true smoking gun of  Mirror Stars and provide a direct window into their interior, allowing measurement of the core temperature and perhaps even aspects of mirror nuclear physics via detailed study of spectral features. 
This is also true for higher $\epsilon \gtrsim 10^{-10}$, where Mirror Stars might appear similar to white dwarfs, providing additional motivation to study them with X-ray observations, see also~\cite{Dessert:2019sgw}. 

The methods we present here and in~\cite{bigpaper} can be readily applied to the signatures of Mirror Stars arising in more realistic hidden sectors, like those arising in Neutral Naturalness~\cite{Chacko:2016hvu,Craig:2016lyx,Chacko:2018vss, MTHastro}.
The optical luminosity scales with the photon kinetic mixing, as well as the size and age of the MS;
 the SM nugget temperature is roughly set by SM ionization energies; and the X-ray signal depends mostly on the  core temperature.
Our more detailed analysis~\cite{bigpaper} also considers mirror-helium-rich Mirror Stars and treats the optically thick regime as well as absorption of optical SM photons in the MS more carefully~\footnote{For $\epsilon \gtrsim 10^{-10}$, the SM nuggets of some benchmark stars considered here become optically thick to bremsstrahlung photons, necessitating different methods for estimating the spectrum shape. There is also some attenuation of the thermal signature by absorption in the MS.}, but this does not affect our conclusions or greatly change the lessons of Figure~\ref{figure:HRplot}.

We have shown that Mirror Star signals are highly distinctive and robust. They  arise in well-motivated theories that may not show up in collider measurements. This makes dedicated searches for Mirror Stars a new frontier in DM detection with completely untapped discovery potential, and an opportunity we cannot afford to miss.

\emph{Acknowledgements:} 
We are grateful to Christopher Matzner and Yoni Kahn for comments on a draft  of this letter.
We also thank 
Christopher Matzner
 Zackaria Chacko, Christopher Dessert, Michael Geller, Bob Holdom, Benjamin Safdi, 
 Kai Schmidt-Hoberg,
 and Yuhsin Tsai
   for helpful conversations. DC would like to especially thank Zackaria Chacko, Michael Geller and Yuhsin Tsai for early discussions on the possibility of mirror stars in MTH models.
 The research of DC and JS was supported by a Discovery Grant from the Natural Sciences and Engineering Research Council of Canada.

\bibliography{References}

\end{document}